\documentclass[12pt,preprint]{aastex}
\usepackage{amssymb}

\shorttitle{Mid-IR Detection of a Hot Core in G29.96-0.02}
\shortauthors{De Buizer et al.}

\begin{document}

\title{Mid-Infrared Detection of a Hot Molecular Core in G29.96-0.02}
\author{James M. De Buizer\altaffilmark{1,2}, Alan Watson\altaffilmark{3},
James T. Radomski\altaffilmark{4}, Robert K. Pi\~{n}a\altaffilmark{4}, and
Charles M. Telesco\altaffilmark{4}}

\begin{abstract}
We present high angular resolution ($\thicksim $0\farcs5) 10 and 18 $\micron$
images of the region around G29.96-0.02 taken from the Gemini North 8-m
telescope using the mid-infrared imager and spectrometer OSCIR. These
observations were centered on the location of a group of water masers, which
delineate the site of a hot molecular core believed to contain an extremely
young, massive star. We report here the direct detection of a hot molecular
core at mid-infrared wavelengths at this location. The size and extent of
the core at 18 $\micron$ appears to be very similar to the morphology as
seen in integrated NH$_{3}$ maps. However, our observations indicate that
the mid-infrared emission may not be exactly coincident with the NH$_{3}$\
emission. 
\end{abstract}


\altaffiltext{1}{Cerro Tololo Inter-American Observatory, National Optical Astronomy
Observatory, Casilla 603, La Serena, Chile.
CTIO is operated by AURA, Inc.\ under contract to the National Science
Foundation.} \altaffiltext{2}{Visiting Astronomer, Gemini North Observatory.}
\altaffiltext{3}{Instituto de Astronom\'{\i}a, Universidad Nacional Aut\'{o}noma de
M\'{e}xico, Apdo. Postal 3-72 (Xangari), 58089 Morelia, Michoac\'{a}n, Mexico.}
\altaffiltext{4}{Department of Astronomy, 211 Space Sciences Research
Building, University of Florida, Gainesville, FL 32601}

\section{Introduction}

Very little is known about the earliest stages of massive star formation. O
and B stars remain embedded in a shroud of obscuring natal material until
they have evolved well past the zero age main sequence. This heavy
extinction makes observations of high mass stars nearly impossible at
wavelength shorter than 1 $\micron$. In fact, little is still known about
high mass star formation, even with the advent of sensitive near-infrared
detector arrays in the recent decade.

Some observations of the earliest stages of massive star formation have come
from deep molecular line imaging at the Very Large Array (VLA) of star
forming regions containing water masers. These radio wavelength searches
were seeking a molecular component to the ultracompact HII (UC HII) regions
associated with young massive stars \citep{CCHWK94}. Owing to the high
resolution and accurate astrometry of the VLA data, these new molecular line
images showed that many water masers are not coincident with the UC HII
regions at all, but instead are coincident with small, non-radio emitting
cores that can be seen in molecular and sub-millimeter emission.

\citet{CCHWK94} observed four sites of UC HII regions in molecular
transitions of NH$_{3}$. They found small structures ($\thicksim $0.1 pc),
with kinetic temperatures greater than 50 and up to 200 K, densities
approximating 10$^{7}$ cm$^{-3}$, and masses of a couple hundred solar
masses. \citet{W95} proposed that these hot molecular cores (or HMCs) could
contain recently formed OB stars which are still undergoing an intense
accretion phase. In this scenario it is believed that the water masers are
excited by the embedded massive stellar sources and exist in their accreting
envelopes, an idea originally proposed for OH masers by \citet{MR68}. In
addition to the lack of observable emission at wavelengths less than 1 $%
\micron$, \citet{W95} also argues that the high mass accretion rates of
massive stars at these early stages could inhibit the onset of an observable
UC HII region.

\citet{CCHWK94} argue that gas and dust would be well mixed in the ammonia
cores and that there would be a high rate of collisions between the dust and
gas. \citet{KW84} estimate that temperature equilibrium between gas and dust
should exist when n$_{H_{2}}>10^{5}$ cm$^{-3}$. Given the
high densities of these cores, it is likely that the gas kinetic temperature
is a fair approximation to the dust temperature. The gas temperatures were
observed by \citet{CCHWK94} to be between 50 and 165 K. They argue that at
these temperatures, the mid-infrared would be a promising wavelength regime
for the discovery of more of these HMCs.

Of the four sources in the sample of \citet{CCHWK94}, a source of great
interest is G29.96-0.02. This is a clear case where the ammonia observations
show a warm compact source, clearly offset ($\thicksim $2$\arcsec$) from a
neighboring UC HII\ region and any extended radio continuum emission. The
estimated core temperature is warm enough to detect at mid-infrared
wavelengths, and the new generation of 8-meter class telescopes allows a
resolution capable of separating the emission of the HMC from the warm dust
in the nearby UC HII region. We present in this paper high resolution
mid-infrared images of the region around G29.96-0.02, and the direct
mid-infrared detection of the HMC at this site.

\section{Observations and Data Reduction}

Observations were obtained at the Gemini North Observatory 8-m telescope on
8 May 2001. The University of Florida mid-infrared camera/spectrometer OSCIR
was used for all observations. OSCIR employs a Rockwell 128x128 pixel Si:As
BIB (blocked impurity band) detector, with a 0\farcs084 pixel$^{-1}$ scale
at Gemini. The total field of view of the array is 11$\arcsec\times $11$%
\arcsec$. Images were taken through two filters, $N$ ($\lambda _{o}$=10.46 $%
\micron$, $\Delta \lambda $=5.1 $\micron$) and $IHW18$ ($\lambda _{o}$=18.06 
$\micron$, $\Delta \lambda $=1.7 $\micron$), centered on the coordinates of
the radio continuum peak position of the UC HII region at R.A.(J2000)=18$%
^{h} $46$^{m}$03\fs93, Decl.(J2000)=-02$\arcdeg$39$\arcmin$21\farcs9 %
\citep{HC96}. Images presented in this paper have on-source exposure times
of 180 seconds, and were taken at an airmass of 1.5. Background subtraction
was achieved during observations via the standard chop-nod technique. Flux
calibration was achieved by observing at a similar airmass the mid-infrared
standard star $\alpha $ Lyr, for which the flux densities were taken to be
38 Jy in the 10 $\micron$ filter and 12 Jy in the 18 $\micron$ filter. A
point spread function (PSF) star was observed in conjunction with these
observations to give an estimate of the spatial resolution of the
observations. The full-width-half-maximum (FWHM) for the PSF star was 0\farcs%
48 at 10 $\micron$ and 0\farcs63 at 18 $\micron$. Subtracting in quadrature
the theoretical diffraction width yields for these observations an
atmospheric seeing and/or telescope guiding contribution of $\thicksim $0%
\farcs40 at both 10 and 18 $\micron$.

During engineering work before these observations, a droplet of glycol
coolant fell onto the dewar window, creating a small circular region of
higher emissivity (and correspondingly, a lower window transmission) in the
lower left (i.e. southwest) quadrant of the array. Though flatfielding did
remove most of the effects of the spot (which is effectively 3 arcseconds in
diameter), the spot in reality may change both the morphology and integrated
flux of any source lying over it. Unfortunately, given the small field of
view of OSCIR\ at Gemini and the large extent of the region being observed,
the southern half of the UC HII region does lie over this spot.

\section{Results and Discussion}

The 10 and 18 $\micron$ images are presented in Figure 1. The UC HII region
is extremely bright in these images, and appears similar in morphology to
the lower resolution mid-infrared images from \citet{BMKAJ96}, and to the 2
cm radio continuum maps of \citet{HC96}. We caution however that the actual
mid-infrared morphology of the UC HII region may be different than presented
in Figure 1 due to the effects of the glycol spot on the window.
Furthermore, the flux of the UC HII region, may be likewise effected. Since
the mid-infrared emission from the UC HII region extends beyond the field of
view of the array, and since the glycol spot may affect the emission we do
see, we do not present here any flux density estimates of the UC HII region
itself. The HMC, on the other hand, was purposely placed on the array as far
as possible from the glycol spot, and therefore the fluxes and morphology of
the HMC presented here are not affected.

\subsection{Background Subtraction and Flux Densities}

There is some difficulty in assessing the flux densities for the HMC because
it lies in the extended dust emission from the UC HII region. However, in
both the 10 and 18 $\micron$ images we were able to subtract out the UC HII
region and background extremely well by fitting them with a 2 dimensional
polynomial surface of fifth order in x and y, excluding a 2$\arcsec\times $ 2%
$\arcsec$ rectangular region around the HMC (Figure 2). The flux densities
for the HMC were found to be 113$\pm 17$ mJy in the 10 $\micron$ filter and
2280$\pm 340$ mJy in the 18 $\micron$ filter. Errors in the flux density
measurements from uncertainty in the standard star flux, atmospheric
variability, and uncertainty in the background subtraction lead to the
quoted $\pm 15\%$ photometric error.

The HMC\ can be seen best at 18 $\micron$, where it is brighter. This large
difference between the 10 and 18 $\micron$ flux densities can be due to two
factors. First, the HMC\ is cool and therefore peaks at wavelengths around
30 $\micron$. Second, models by \citet{OLA99} of massive star formation via
spherical accretion of a free-falling envelope of gas and dust show that
silicate absorption is a prominent spectral feature at 10 $\micron$ during
early stages of massive stellar evolution. We therefore would expect all
HMCs to be more readily observed in the mid-infrared at wavelengths greater
than 14 $\micron$ due to the rising SED and the absence of significant
absorption at these longer wavelengths.

However in this particular case, the measured mid-infrared flux densities
for the HMC in G29.96-0.02 yield a derived color temperature of 118 K under
the assumption of optically thick emission, and 105 K in the optically thin
case. The presence of cooler overlying absorbing dust would require yet
higher temperatures. All of these temperature estimates are significantly
warmer than the temperature of 85 K estimated by Cesaroni et al. (1994) from
observations of NH$_{3}$. The optically thin temperature is closest to the
temperature derived from the ammonia observations, and may indicate that the
mid-infrared emission is therefore optically thin. Moreover, this
temperature difference is unlikely to be a result of differences between the
local temperatures of the dust and gas, as Cesaroni et al. (1994) noted that
these should be well-coupled. A possible explanation for this difference
is that the optical depth unity surface of NH$_{3}$ may be at a slightly 
larger radius in the HMC than the optical depth unity surface at 10 $\micron$.
Furthermore, we could be averaging over regions with different temperatures,
possibly because of the presence of a radial temperature gradient or
unresolved multiple sources, and that warmer regions contribute more strongly 
to the mid-infrared. 

\subsection{Hot Core Morphology}

At both 10 and 18 $\micron$ the core is resolved, and is elongated in the
E-W direction (Figure 2). The overall size and morphology of the HMC at 18 $%
\micron$ is similar to the integrated NH$_{3}$(4,4) maps of \citet{CHWC98}.
From gas kinematic studies, it was found that there is a velocity gradient
along this axis of elongation in the molecular lines of NH$_{3}$(4,4) %
\citep{CHWC98} and SiO(2$-$1) $v=0$ \citep{MTCW01}. The combination of
elongation of the molecular material and this velocity gradient has lead to
the suggestion that this is a rotating disk around an accreting massive
protostar.

The elongation of the mid-infrared source is very similar in both 10 and 18 $%
\micron$. The Gaussian fits to the HMC yield a position angle of 94$\arcdeg$
and a major-to-minor FWHM ratio of 1.26 for the 10 $\micron$ image and 1.25
for 18 $\micron$. The observed FWHM of the major axis of elongation at 10 $%
\micron$ is 1\farcs33 and at 18 $\micron$ it is 1\farcs42. The deconvolved
source size obtained by subtracting the PSF FWHM in quadrature from the
observed FWHM for the HMC is 1\farcs23 at 10 $\micron$ and 1\farcs27 at 18 $%
\micron$. This corresponds to physical diameters of 7380 and 7620 AU,
respectively, using a distance of 6.0 kpc \citep{PMB99}.

\subsection{Astrometry}

The absolute positions of the mid-infrared sources are unknown. No attempt
was made to achieve accurate astrometry at the telescope due to time
constraints. However, we detected the mid-infrared component of the UC HII
region and the HMC, both of which have radio positions known to high
absolute accuracy. In spite of this, there is still an ambiguity in
establishing the correct astrometry of the sources. This stems from the fact
that the peak of the UC HII region and the peak of the HMC are farther apart
in the mid-infrared (2\farcs6) than in the radio (2\farcs1). This yields two
possible astrometric scenarios.

\subsubsection{Aligning the emission from the hot core}

Because the morphology of the HMC at 18 $\micron$ is similar to the ammonia
morphology in size, extent, and shape, we might assume that the emission is
coming from the same source. Astrometry could be determined by registering
the peaks of the HMC in the 18 $\micron$ map and the ammonia map of %
\citet{CHWC98}. In this astrometric scenario, the easternmost water masers
of \citet{HC96} would be coincident with the ammonia and mid-infrared peaks.
The rest of the water masers to the west of the peak, as well as the
methanol masers, would then appear to be coming from a region traced by the
outer contours of the mid-infrared and ammonia emission.

The problem with this alignment is that it requires that the arc of infrared
emission from the UC HII region lie to the east of the arc of radio
continuum emission. As the ionizing star is located within the arc %
\citep{Alan97}, this requires that the mid-infrared emission arises closer
to the ionizing star than the radio continuum emission. Given the similarity
of the morphology of the arc in the mid-infrared and radio, and given that
both are generated at locations where stellar photons encounter dense gas,
we would expect both arcs to closely coincide.

\subsubsection{Aligning the emission from the UC HII region}

The more plausible scenario is to register the mid-infrared and radio images
by aligning the peaks of the UC HII region at both wavelengths. In this
case, however, we have the mid-infrared emission peak of the HMC offset to
the southwest 3000 AU (0\farcs5) of the HMC peak seen in ammonia emission.
This offset between emission at molecular wavelengths and the mid-infrared
is similar to that seen by \citet{KPBAJ92} for the HMC in W3(OH) which are
offset by about 3500 AU. We show this astrometric scenario in Figure 3 with
the locations of the water \citep{HC96} and methanol \citep{MCB2001} masers
overplotted. The postions of these masers are known to a high astrometric
accuracy with respect to the ammonia emission ($\lesssim 0.3\arcsec$).

We could conjecture in this scenario that the offset between the ammonia and
mid-infrared peaks may be due to optical depth effects. The core might be so
dense that mid-infrared emission cannot escape here, and instead the
mid-infrared emission that we are seeing is apparently tracing hot dust in a
less embedded environment to the northeast. However, as metioned earlier,
the color temperatures derived from the mid-infrared flux densities seem to
indicate that the mid-infrared emission is optically thin. Furthermore, we
noted that the difference between the mid-infrared derived color temperature
and temperature derived from ammonia observations may be due to the presence
of multiple sources. A second conjecture is that there may be two sources in
close proximity to each other, making the source appear to be extended at
both wavelengths. The observed extended emission of both the dust and
ammonia might overlap, but the ammonia emission peaks with the easternmost
water masers, while the mid-infrared dust emission appears to wrap around
the ammonia peak. These two sources could therefore be in slightly different
evolutionary stages, with the eastern source being extremely young and
highly embedded and the western source being more evolved and therefore less
embedded.

Interestingly, given this alignment, the methanol masers are more closely
associated with the mid-infrared peak than the ammonia peak. This is
consistent with the models of \citet{SD94} and \citet{SCG97} that indicate
methanol masers are pumped by mid-infrared photons.

\subsection{Millimeter Observations and SED Modeling}

Apart from the mid-infrared continuum observations presented here, this HMC
has also been observed at two other continuum wavelengths, namely 1 and 3 mm
(Maxia et al. 2001). If the mid-infrared emission indeed comes from the same
location as the 1 and 3 mm emission, one could model the SED from this
source and determine physical parameters, such as accretion rates and
luminosities \citep{OLA99}. Even if we believe that there is a single HMC
responsible for the emission at all wavelengths, there appears to be
problems with the 1 and 3 mm flux densities of Maxia et al. (2001). The
slope of the SED in the Rayleigh-Jeans portion of the distribution is
predominantly determined by the dust opacity at these wavelengths. Most
models adopt a power law for the dust opacity of the form, $\kappa _{\lambda
}\propto \lambda ^{-\beta }$, where $1\leq \beta \leq 2$ for $\lambda \geq
200$ $\micron$. Given the flux density values of Maxia et al. (2001), the
value for $\beta $ would be negative, which seems unlikely. It seems
plausible that the large beam size of the millimeter observations (HPBW$%
_{average}$ $\simeq 4\arcsec$) of Maxia et al. (2001) may have led to an
inaccurate subtraction of the millimeter emission of the UC HII region from
that of the HMC. Futhermore, observations at 3 mm are more prone to 
contamination by free-free emission from the nearby UC HII region than at 1 
mm. At present therefore, we are only able to constrain the
Wien side of the SED using our two mid-infrared flux density measurements,
and the Rayleigh-Jeans side by upper limits from other observations. These
types of loose constraints to the modeling cannot estimate physical
parameters of the HMC with accuracy. Better sampling of the SED is needed
from the mid-infrared to the millimeter.

If there are two HMCs here, the millimeter fluxes from these two source are
blended in the flux densities quoted in Maxia et al. (2001) because of
insufficient spatial resolution. In order to determine whether this site
contains more than one HMC, and whether a more evolved source is responsible
for the mid-infrared emission near the ammonia core, higher spatial
resolution millimeter or submillimeter observations (e.g. ALMA) are required.

\section{Conclusions}

We have directly detected the hot molecular core of G29.96-0.02 at
mid-infrared wavelengths. The size and morphology of the HMC in the
mid-infrared is very similar to its appearance in the integrated NH$_{3}$%
(4,4) maps of \citet{CHWC98}. The mid-infrared emission is elongated E-W, as
is the ammonia emission. It is likely, however, that more than one embedded
stellar source may exist at this location or that the emission in the
mid-infrared and the ammonia emission may not be coming from the same
location in the HMC. Higher resolution observations ranging between the
mid-infrared and millimeter wavelengths are needed to investigate the
possible duplicity of the HMC or to perform any accurate modeling of the HMC
SED.

\acknowledgments We wish to thank Mayra Osorio for her advice and input, as
well as her effort in attempting some preliminary SED models. This paper is
based on observations obtained with the mid-infrared camera OSCIR, developed
by the University of Florida with support from the National Aeronautics and
Space Administration, and operated jointly by Gemini and the University of
Florida Infrared Astrophysics Group. Gemini Observatory is operated by the
Association of Universities for Research in Astronomy, Inc., under a
cooperative agreement with the NSF on behalf of the Gemini partnership: the
National Science Foundation (United States), the Particle Physics and
Astronomy Research Council (United Kingdom), the National Research Council
(Canada), CONICYT (Chile), the Australian Research Council (Australia), CNPq
(Brazil) and CONICET (Argentina).

\clearpage

\begin{figure}[tbp]
\epsscale{1.0} \plotone{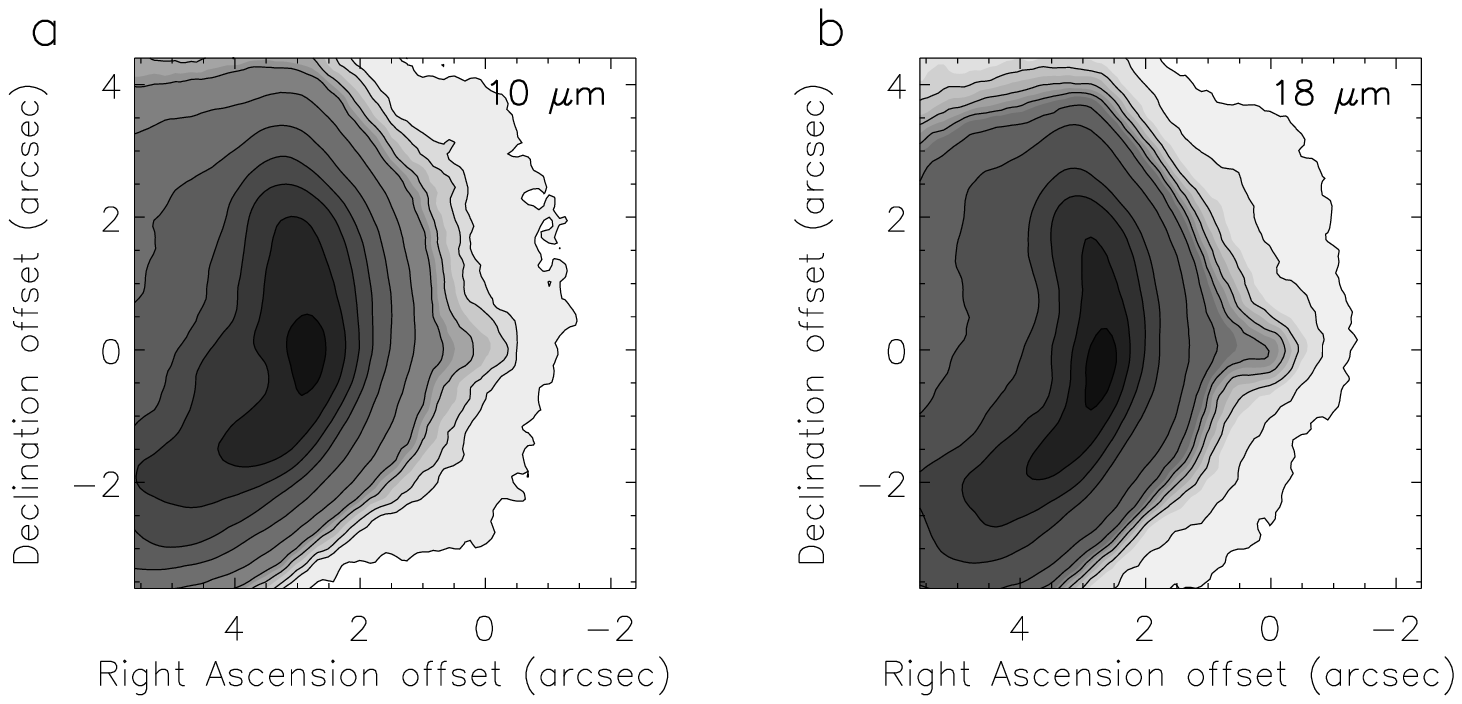} 
\figcaption[G29fig1.eps]{Filled countour maps of G29.96-0.02.  The arc-shaped UC HII region 
dominates the mid-infrared emission, however the hot core can be seen as a ``bump'' approximately 
2$\arcsec$ west of the UC HII peak. Panel (a) shows the 
10 $\micron$ contours at 1, 2, 3, 4, 5, 8, 15, 24, 36, 54, and 84\% of the peak flux 
density of 8.2 Jy arsec$^{-2}$. Panel (b) shows the 18 $\micron$ contours at 2, 3, 5, 6, 8, 10, 15, 
30, 44, 66, and 88\% the peak flux density of 38.9 Jy arsec$^{-2}$.}
\label{fig 1}
\end{figure}

\clearpage

\begin{figure}[tbp]
\epsscale{1.0} \plotone{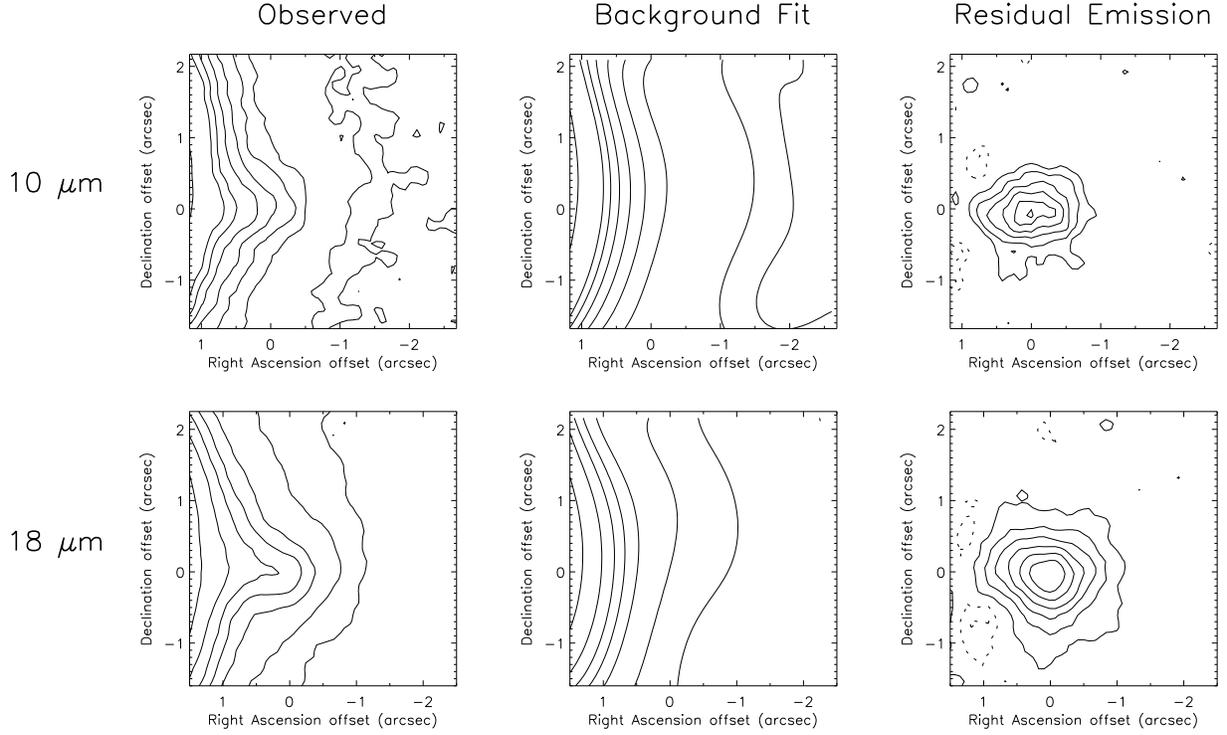} 
\figcaption[G29fig2.eps]{Background- and UC HII-subtracted maps. A two-dimensional 
polynomial surface was fit to the background and UC HII region (excluding a rectangular 2$\arcsec$ 
by 2$\arcsec$ region around the HMC) and subtracted from 
the original image. Because the UC HII region peak is hard to fit in this manner, cropped 
versions of the observed images were used. The origin is the same as for Figure 1. The 
top panels show, from left to right, the cropped 10 $\micron$ image, the background 
polynomial fit of 5th order in x and y, and the hot core in the residual frame. The 
contours shown are -25, -16, 16, 33, 56, 67, 83, and 93\% of the peak flux density of 
170 mJy arsec$^{-2}$. The bottom three panels show the same for the  18 $\micron$ data. The contours 
shown are -16, -8, 8, 25, 42, 59, 76, and 93\% of the peak flux density of 1.9 Jy arsec$^{-2}$.}
\label{fig 2}
\end{figure}

\clearpage

\begin{figure}[tbp]
\epsscale{1.0} \plotone{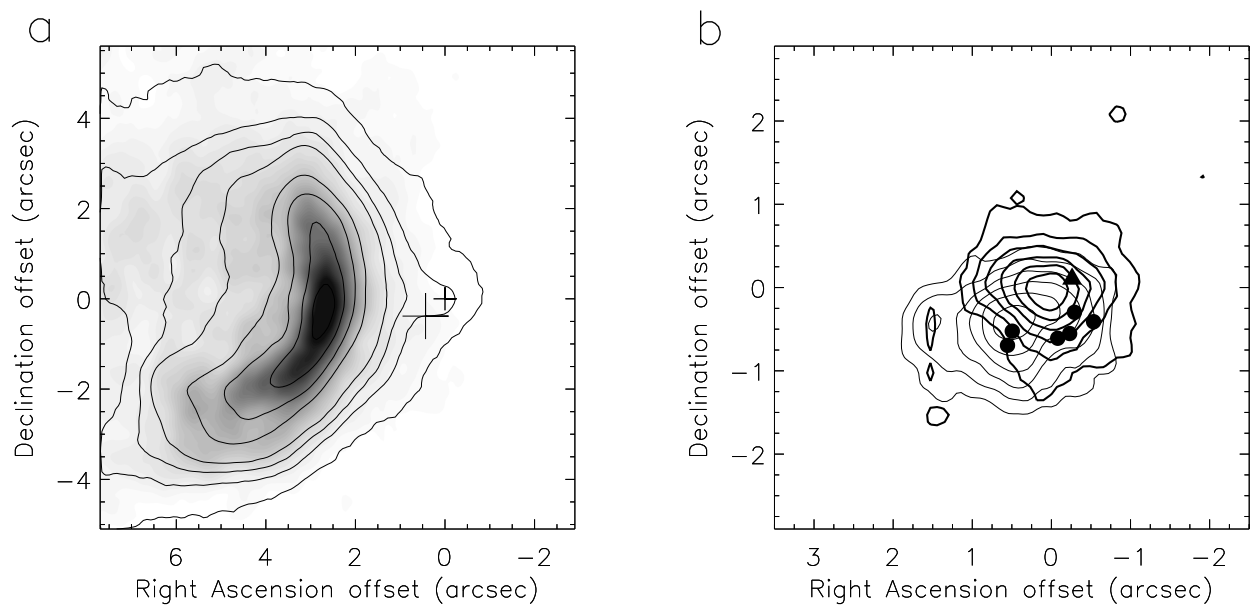} 
\figcaption[G29fig3.eps]{Registration of the mid-infrared data with the radio data by means 
of the UC HII morphology. Panel (a) shows a 
grayscale image of the 24 GHz continuum observations of Cesaroni et al. (1998). 
The overlaying contours are from the 18 $\micron$ mid-infrared data presented here. Good 
morphological agreement is found between the two wavelengths adding credibility 
to the astrometry. The large, thin cross marks the position of the ammonia peak 
of the hot core from Cesaroni et al. (1998). The small, thick cross marks the peak in 
the mid-infrared dust emission from the hot core. Panel (b) shows a close up view 
centered on the mid-infrared peak of the hot core. Apparent in this figure are the offset 
peaks of the mid-infrared (thick contours) and ammonia emission (thin contours), possibly 
due to optical depth effects, or perhaps showing there are two 
different embedded objects here. Filled black circles mark the locations 
of the water masers from Hofner and Churchwell (1996), and the filled triangle the 
location of the methanol maser group from Minier et al. (2001).} %
\label{fig 3}
\end{figure}

\end{document}